\documentclass[aps,prl,twocolumn,groupedaddress,floatfix,showpacs,keywords]{revtex4}

\usepackage{graphicx}
\usepackage{amssymb}

\begin{document}

\title{Signatures of glass formation in a fluidized bed of hard spheres}
\author{Daniel I. Goldman}
\email[]{digoldma@berkeley.edu}
\altaffiliation{Present address: Department of Integrative Biology,
                The University of California at Berkeley,
                Berkeley, CA 94720-3140}
\author{Harry L. Swinney}
\affiliation{Center for Nonlinear Dynamics and Department of
Physics, The University of Texas at Austin, Austin, TX 78712}

\date{\today}

\begin{abstract}

We demonstrate that a fluidized bed of hard spheres during
defluidization displays properties associated with formation of a
glass. The final state is rate dependent, and as this state is
approached, the bed exhibits heterogeneity with increasing time and
length scales. The formation of a glass results in the loss of
fluidization and an arrest of macroscopic particle motion. Microscopic motion persists in this state, but the bed can be jammed by application of a small increase in flow rate. Thus a fluidized bed can serve as a test system for studies of glass formation and jamming.

\end{abstract}


\maketitle
Despite a century of study, there is no unified theory for the formation of a glass state from a liquid cooled below its freezing point (supercooled)~\cite{ediAang,debAsti}. However, some signatures of the glass transition have been identified. One is the dependence of the volume of the glass on the cooling rate: the slower a liquid is cooled, the greater density it achieves when it forms a glass. Another signature is a rapid increase in a characteristic relaxation time (or viscosity) of the liquid during cooling as the glass transition temperature $T_g$ is approached. In colloidal and hard sphere systems the transition is reached by increasing volume fraction $\phi$ rather than decreasing temperature; the transition occurs at $\phi_g \approx 0.58$ ~\cite{pusAvanPRL, vanAund94, speedy98}.

In   liquids~\cite{edigerreview,traAwilshortref,qiuAedi,berthier05shortref}, colloids ~\cite{berthier05shortref,weeAcroshortref}, and simulations~\cite{glotzerreview,dolAheu98,berAbou,garAcha} it has been observed that as the glass transition is approached, the system dynamics becomes increasingly heterogeneous. Both the size of spatially correlated regions and the time scale for rearrangement of these regions increase rapidly near the glass transition.

We examine a fluidized bed~\cite{jacksonbook,sundarreview}, which is a vertical column of particles with an upward flow of fluid that maintains the particles in motion. As the fluid flow rate $Q$ is
decreased, this non-equilibrium~\cite{marty05} system exhibits
characteristic signatures of the equilibrium glass transition. We find that two different volume fractions are important in the loss of fluidization: $\phi_g = 0.582$, where bulk quantities
change, and $\phi_a = 0.594$, where $\phi$ becomes nearly
independent of $Q$ and only small scale motion persists; this motion
can be stopped by a slight increase in flow rate, jamming the
system. Our observed $\phi_g$ is close to the volume fraction at which a glass transition is observed in other hard sphere
systems~\cite{weeAcroshortref,pusAvanPRL,vanAund94,speedy98}. The
observed $\phi_a$ is close to the value obtained for slow
defluidization of particles in a gas fluidized bed~\cite{ojhAmen}, and close to the volume fraction where the system has the maximal number of statistically independent regions~\cite{schAgol}.

\begin{figure}[h!tb] \begin{center} \includegraphics[width=3in]{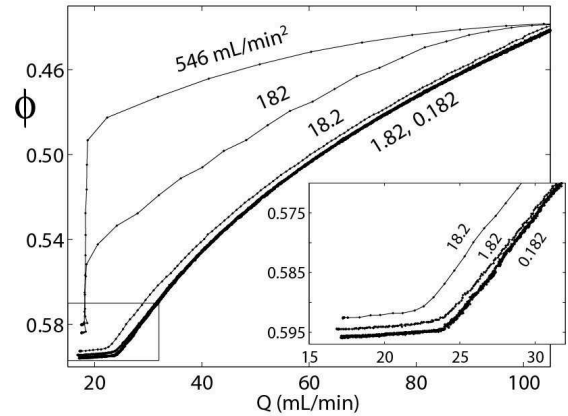} \caption{\label{figure1} The volume fraction achieved during defluidization depends on the ramp rate (cf. values next to the curves), just as in the glass transition in colloids and in hard sphere models. In the top two curves the ramping is so rapid that the final state is achieved through sedimentation. Inset: magnification of the region near the arrest transition at $\phi_a=0.594$; only the three slowest ramp rates are shown.} \end{center}
\end{figure}

{\it Experiment ---} Water flows upward at a volume flow rate $Q$ through a vertical column of glass spheres in a square bore glass tube of cross sectional area $A=5.81~\mbox{cm}^2$. There are $4 \times 10^6$ glass spheres with diameter $d=250 \pm 8~\mu$m and density $\rho_p=2.47~\mbox{g/}\mbox{cm}^3$. Flow rate fluctuations are smaller than $0.3 \%$. To obtain uniform flow, fluid passes into the bottom of the column through a nylon mesh ($5~\mu$m weave, open area $0.75 \%$, Nitex mesh, Sefar America). The average height $h$ of the bed top surface above the distributor is measured to determine the average bed solid volume fraction, $\phi=M_p / \rho_p A h$, where $M_p$ is the total mass of the particles. We measure the fluid pressure drop $\Delta P$ from the bottom to the top of the bed; values are normalized by the buoyant weight of the grains. The dynamics of the bed are studied by imaging the side of the bed and by using Diffusing Wave Spectroscopy (DWS)~\cite{weitzbrown} to probe the interior of the bed.

\begin{figure}[h!tb]
\begin{center} \includegraphics[width=2.5in]{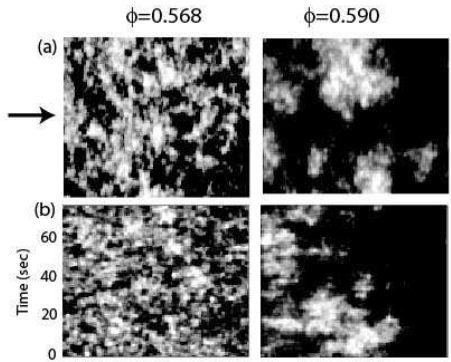} 
\caption{\label{figure2} (a) Dynamical heterogeneity: as the glass
state is approached, transiently immobile regions (black) and
transiently mobile regions (white) grow in size.  Each panel is the
difference between two images ($1.5 \times 1.5~\mbox{cm}^2$) of the
side of the bed taken ${\Delta T} = 0.25$ s apart. (b) Raster scans
for the row of pixels indicated by an arrow in (a) show that the
lifetimes of the correlated regions increases as the transition is
approached. The flow rates for $\phi=0.568$ and $\phi=0.590$ were
respectively 33.02 and 26.19 $\mbox{mL}/\mbox{min}$.  In this figure
and in Fig. 3 the data were obtained at $\phi$ values reached by
decreasing $Q$ at the slowest ramp rate in Fig.~\ref{figure1}.}
\end{center}
\end{figure}

\begin{figure}[h!tb]
\begin{center}
\includegraphics[height=5.65in]{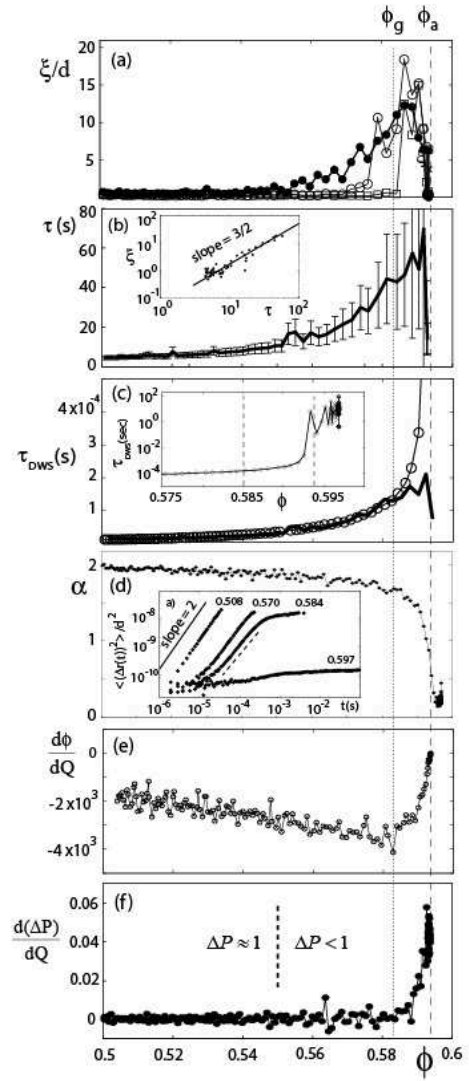} \caption{\label{megafigure} Measurements on a fluidized bed reveal two transitions: a transition in bulk properties at $\phi_g=0.582 \pm 0.004$ and an arrest transition at $\phi_a=0.594 \pm 0.003$~\cite{uncertaintynote}. (a) An increase near $\phi_g$ in the spatial correlation length $\xi$ of mobile regions (Fig. 2(a)), and a decrease in $\xi$ beyond $\phi_g$ to $\xi=0$ at $\phi_a$. ($\xi$ is shown for $\Delta T$=0.25 sec ($\bullet$), 2.8 sec ($\circ$), and 50 sec ($\square$).) (b) An increase in the average correlation time in bed images (Fig 2(b)). The inset shows a power law dependence of $\xi$ on $\tau$ near $\phi_g$. (c) An increase near $\phi_g$ in the decorrelation time $\tau_{\tiny \mbox{DWS}}$ from the intensity autocorrelation function (1/e point) of transmitted laser light~\cite{weitzbrown}. The bold solid curve, the correlation time from (b) scaled by $3 \times 10^5$, begins to deviate from $\tau_{\tiny \mbox{DWS}}$ at $\phi_g$. The inset shows that at the arrest transition, $\tau_{\tiny \mbox{DWS}}$ no longer increases. (d) A decrease beyond $\phi_g$ in the exponent $\alpha$ for the Mean Square Displacement (MSD) for intermediate times indicated by the dashed line in the inset, which shows the time dependence of MSD. (e) An inflection in $d \phi / dQ$ at $\phi_g$ and a divergence in $d \phi / dQ$ as $\phi_a$ is approached. (f) An increase in $d(\Delta P)/dQ$ beyond $\phi_g$ and a peak at $\phi_a$, where $d(\Delta P)/dQ$ becomes independent of $Q$.}
\end{center}
\end{figure}

{\it Signatures of glass formation ---} The dependence of $\phi$ on the ramping rate (Fig.~\ref{figure1}) is similar to the dependence of glass volume on cooling rate in supercooled liquids that form glasses~\cite{ediAang}. In a fluidized bed for sufficiently large $Q$, the grains are mobile~\cite{menAdur97a} and $\Delta P \cong 1$. However, as $Q$ decreases, particle mobility decreases and $\phi$ increases until the defluidization transition is reached at $\phi_a$ (the knee of the curve in the inset of Fig. 1); $\phi_a$ decreases only $1 \%$ as the ``cooling rate" $dQ/dt$ is decreased by more than two orders of magnitude. 

Another signature of glass formation is dynamical heterogeneity. As in other glass-forming systems~\cite{glotzerreview, edigerreview, garAcha, weeAcroshortref, qiuAedi}, the size of the mobile and immobile regions in the fluidized bed increases as $\phi$ increases~\cite{weeAcroshortref}, as shown by the images in Fig.~\ref{figure2}(a) and by the correlation range $\xi$ in Fig.~\ref{megafigure}(a), where $\xi$ is given by the $1/e$ point of an azimuthal average of 2D autocorrelation of difference images separated by time $\Delta T$. The heterogeneities observed both at the top and the side of the bed are similar to volcanos and channels
observed in gas fluidized beds~\cite{volcandchannelsnote}.

The decrease in $\xi$ near $\phi_g$ is similar to the decrease in length scale of cooperative regions observed in colloid experiments~\cite{weeAcroshortref} and lattice model simulations~\cite{panAgar}. The decrease in $\xi$ that we observe arises because the dynamics has slowed enough near $\phi_g$ so that no discernible motion occurs during the measurement time $\Delta T$. Motion can be observed for increased $\Delta T$, but then $\xi$ remains close to zero until $\phi \approx \phi_g$ (cf. Fig.~\ref{megafigure}(a)~\cite{uncertaintynote}); most particles in the bed have moved during this $\Delta T$ and thus the correlation measurement picks up only the minimum observable motion.  The lifetime $\tau$ of mobile and immobile regions increases as $\phi$ increases, as shown in Fig.~\ref{figure2}(b) and in Fig.~\ref{megafigure}(b), where $\tau$ is obtained by averaging over all pixels (i,j) (for each pixel the time $\tau_{i,j}$ is given by the $1/e$ point of the time autocorrelation function of the intensity). The standard deviation of the distribution of decay times (denoted by vertical bars in Fig.~\ref{megafigure}(b)) also increases as $\phi$ increases; similar behavior has been found in a lattice model~\cite{panAgar}. The slowing of the dynamics is similar to other glass forming systems~\cite{ediAang}. We find that the length and time scales for $\phi<\phi_g$ are linked by a power law, as shown in the inset of Fig.~\ref{megafigure}(b).  A power law was also obtained in a study of a lattice model~\cite{panAgar}, but with an exponent value 1/4 rather than the 3/2 given by our observations. In contrast to~\cite{panAgar} the lengthscale plotted in  Fig.~\ref{megafigure}(b) is obtained for fixed $\Delta T=0.25$ sec. 

Additional signatures of glass formation are revealed by the bed behavior on short time scales. Measurements of the intensity-intensity correlation function $g^{(2)}(t)$ of light multiply scattered as it travels through through the bed~\cite{lstarcomment} yield the time scale $\tau_{\tiny \mbox{DWS}}$ (Fig.~\ref{megafigure}(c)), which increases until $\phi=\phi_g$ with the {\em {same}} functional form as the decorrelation time of image pixel intensity $\tau$ (Fig.~\ref{megafigure}(b)). The correspondence in the behavior for times differing by a factor of $\sim 10^5$ indicates that, as in other glass formers~\cite{ediAang}, the macroscopic and microscopic dynamics behave the same prior to the glass transition and become decoupled as the system forms a glass.

Information on particle motion at small length scales was obtained using Diffusing Wave Spectroscopy (DWS) theory~\cite{weitzbrown} to invert the autocorrelation curves and obtain $\langle \Delta r(t)^2 \rangle$, the mean square displacement (MSD) of the grains at times so short that particles moved only $\sim 0.01\%$ of their diameter. For $\phi \ll \phi_g$, the short time dynamics are well described by $\langle \Delta r(t)^2 \rangle \propto t^\alpha$ with $\alpha \approx 2$ (Fig.~\ref{megafigure}(d) inset), indicating that all grains undergo ballistic motion between collisions~\cite{menAdur97a,menAdur97b}. For $\phi \gtrapprox 0.54$, as $\phi$ approaches $\phi_g$, the curves develop a region with $\alpha <2$ at short times, indicating that at these times the particles remain in contact with neighbors. However, at larger times, $\alpha \approx 2$ (cf. inset of Fig.~\ref{megafigure}(d)), indicating that the particles move ballistically. As $\phi$ approaches $\phi_g$, the MSD develops a plateau region at longer times, indicating the particles are caged; caging has been observed in hard sphere systems that approach the glass state~\cite{weeAcroshortref,dolAheu98,vanAund94}. However, the plateau observed in our MSD measurements indicates a caging length of order $10^{-4}d$, three orders of magnitude smaller than what is observed in these systems. For $\phi > \phi_g$, the slope at intermediate times decreases, indicating that particles are no longer able to break from their cages, and $\alpha$ rapidly decreases (Fig.~\ref{megafigure}(d)). For $\phi > \phi_a$, all particles are immobile on macroscopic length scales (Fig~\ref{megafigure}(a)-(c)).

The bulk properties of the bed change at $\phi_g$: $d \phi / dQ$ reaches a minimum (Fig.~\ref{megafigure}(e)), and the pressure drop $\Delta P$ across the bed begins to decrease rapidly, dropping well below the buoyant weight of the bed~\cite{valAcas}. This is shown by the derivative $d(\Delta P)/dQ$, which increases at $\phi_g$ until the increase is arrested at $\phi_a$ (Fig.~\ref{megafigure}(f)), beyond which the bed has defluidized and $\Delta P$ becomes linear with $Q$, following Darcy's law~\cite{batchelor67}. We propose that the minimum in $d \phi /dQ$ and the increase of $d(\Delta P)/dQ$ beyond $\phi_g$ are further signatures of glass formation: once the bed begins to behave like a glass close to $\phi_g$, the necessity for increasingly cooperative particle motion~\cite{weeAcroshortref} makes it unable to continue to pack sufficiently (increase $\phi$) in response to changes in $Q$ as it would in the fluidized state where $\Delta P$ is maintained close to unity. Since the formation of a glass triggers the rapid drop in $\Delta P$, which leads to the final macroscopic arrest of the bed at $\phi_a$, we conclude that defluidization is a consequence of the hard sphere glass transition. This explains why $\phi_a$ seen in our water fluidized bed and the gas fluidized bed experiments of Ojha et al.~\cite{ojhAmen} is independent of particle size, container aspect ratio, and fluidizing medium.

\begin{figure}[h!tb] \begin{center} \includegraphics[width=2.75in]{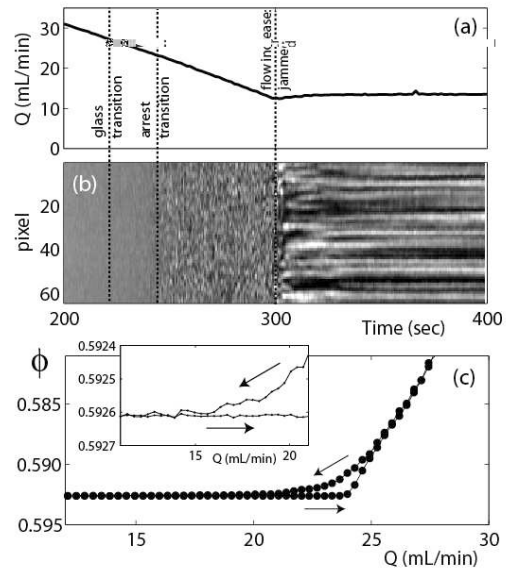} \caption{\label{figure4} Microscopic motion in the glass state is halted by a slight increase in $Q$, jamming the system. (a) $Q$ is decreased (at $9.6~\mbox{mL}/\mbox{min}^2$) to $12.5$ mL/min, at which point it is increased by $1$ mL/min (indicated by the rightmost dashed line). The glass state (when $\Delta P<1$) is achieved at $Q=24$ mL/min, indicated by the leftmost dashed line. (b) The speckle field, measured (30 ms exposure) along a line with each pixel corresponding to a single coherence area, becomes time independent after jamming occurs. (c) Hysteresis in $\phi$ for increasing and decreasing $Q$ after an increase in $Q$ at $Q \approx 8$ mL/min. The hysteresis disappears for $\phi > \phi_g$.} \end{center} \end{figure}

{\it Jamming ---} Even in the arrested state, microscopic motion persists, as shown in the time evolution of images of a line of scattered light~\cite{wonAwil} (Fig.~\ref{figure4}(b)).  The intensity of each pixel fluctuates even for $\phi > \phi_a$, indicating that microscopic motion persists. However, for $\phi > \phi_g$ all microscopic motion can be jammed (stopped) by slightly increasing $Q$ (for the conditions of our experiment, DWS can detect motion of only 1 nm for any scattering particle). The jamming presumably establishes the stress backbone of the system~\cite{oheAlan01} and explains the hysteresis seen in Fig.~\ref{figure4}(c). Once the system is jammed, increases and decreases in $Q$ below the onset of fluidization~\cite{jacksonbook} do not change $\phi$. Such jamming has been associated with glass states~\cite{liuAnag98,oheAlan01,oheAsil} and has been studied in fluidized beds~\cite{valAqui04}.

{\em Conclusions ---} We have shown that a fluidized bed exhibits essential features of glass formation: rate dependence on final state, dynamical heterogeneity, rapid increase in time scale, and jamming. Further, the dynamics of the hard sphere glass formation control the defluidization of the bed: beyond $\phi_g$ the system cannot pack sufficiently to accommodate changes in flow and must therefore arrest at $\phi_a>\phi_g$ with $\phi_a$ independent of particle size, container aspect ratio ~\cite{ojhAmen}, and fluidizing medium. The volume fraction transition values depend weakly on surface properties~\cite{schAgol} and strongly on cohesive effects~\cite{valAcas}. Finally, we speculate that the Random Loose Packed volume fraction, $\phi_{\tiny \mbox{RLP}} \approx 0.56$~\cite{onoAlin}, plays a role in the onset of heterogeneity, similar to the concept of onset transition observed prior to the glass transition~\cite{schAsas00}.

A fluidized bed is a simple system that allows fine control, so it is an ideal system for studying glass and jamming transitions and for informing theory~\cite{bazant}. Further, understanding glass behavior in a fluidized bed can inform fluidized bed design, which is important in many industrial applications~\cite{jacksonbook}.

\begin{acknowledgments}
We thank M. Shattuck for experimental assistance and discussion, and
M. Schr\"oter, D. Chandler, J. Garrahan, A. Pan, and E. Weeks for
helpful discussions. This work was supported by Robert A. Welch
Foundation and the Office of Basic Energy Sciences of the
Department of Energy.

\end{acknowledgments}


\begin{thebibliography}{36}
\bibitem[{unc()}]{uncertaintynote}
\bibinfo{note}{The uncertainty is $\phi_g$ has been obtained by fitting the
  minimum in Fig.~\ref{megafigure}(e) to a parabola, by determining the
  intersection of fit lines below and above $\phi_g$ in
  Fig.~\ref{megafigure}(f), from the point of separation of the two curves in
  Fig.~\ref{megafigure}(c), and uncertainties in $\rho_p$ and $A$, as well as
  rate dependent effects. The uncertainity in $\phi_a$ arrises from uncertainty
  in $\rho_p$ and $A$.}
\bibitem[{lst()}]{lstarcomment}
\bibinfo{note}{We determine the scattering length $l^{*}$ over which photon
  orientation is uncorrelated using a high speed camera to estimate the
  collision velocity of particles near the side of the bed. We then compute
  $l^{*}$ for each $\phi$ by taking the ratio of the $\phi$ values. For all
  $\phi$ we find $L/l^{*} \gtrapprox 10$.}
\end{thebibliography}
\end{document}